\documentclass[12pt,a4paper]{article}
\usepackage[a4paper]{geometry}
\usepackage[utf8]{inputenc}
\usepackage[english]{babel}
\usepackage{amsmath}
\usepackage{amstext}
\usepackage{amscd}
\usepackage{amsfonts}
\usepackage{amssymb}
\usepackage{mathrsfs}

\newtheorem{proof*}{Proof}

\usepackage[pagebackref,draft=false]{hyperref}
\hypersetup{colorlinks,
linkcolor=myrefcolor,
citecolor=mycitecolor,
urlcolor=myurlcolor}

\usepackage[capitalize]{cleveref}
\usepackage{cite}
\usepackage{caption}
\usepackage{etaremune}

\usepackage{xcolor}
\definecolor{myurlcolor}{rgb}{0,0,0.4}
\definecolor{mycitecolor}{rgb}{0,0.5,0}
\definecolor{myrefcolor}{rgb}{0.5,0,0}
\usepackage{graphicx}
\usepackage{tikz}
\usepackage{tikz-cd}
\usetikzlibrary{cd}
\usetikzlibrary{decorations.pathmorphing,shapes}

\usepackage{etoolbox}
\usepackage{enumitem} 
\usepackage[]{latexsym}
\usepackage{braket}
\usepackage{caption}
\usepackage[utf8]{inputenx}
\usepackage[T1]{fontenc}
\usepackage{lmodern}
\usepackage{textcomp}
\usepackage{microtype}
\usepackage{totcount}
\usepackage{blindtext}

\newcommand{\be}{\begin{equation}}
\newcommand{\ee}{\end{equation}}

\title{Some remarks on the notion of transitions}
\date{}

\author{F. M. Ciaglia $^{1,3}$ \href{https://orcid.org/0000-0002-8987-1181}{\includegraphics[scale=0.7]{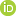}}, F. Di Cosmo$^{1,2,4}$ \href{https://orcid.org/0000-0003-0256-5913}{\includegraphics[scale=0.7]{ORCID.png}}\\
\footnotesize{$^{1}$\textit{Depto. de Matem\'aticas, Univ. Carlos III de Madrid, Legan\'es, Madrid, Spain}} \\
\footnotesize{$^{2}$\textit{ ICMAT, Instituto de Ciencias Matem\'{a}ticas (CSIC-UAM-UC3M-UCM)}}\\
\footnotesize{$^{3}$\textit{ e-mail: \texttt{fciaglia[at]math.uc3m.es}}} \\
\footnotesize{$^{4}$\textit{ e-mail: \texttt{fcosmo[at]math.uc3m.es}}} 
}

\begin{document}

\maketitle
\begin{abstract}
In this paper some reflections on the concept of transition are presented: groupoids are introduced as models for the construction of a ``generalized logic'' whose basic statements involve pairs of propositions which can be conditioned. In this sense, we could distinguish between classical probability theory where propositions can be conditioned if they have a non-zero intersection, from cases  where ``non-local'' conditioning are allowed. The algebraic and geometrical properties of groupoids can be exploited to construct models of such non-local description.

\end{abstract}
\section{Introduction}
The origin of information theory dates back to the work by Shannon on the mathematical theory of communication \cite{Shannon}. His aim was to introduce a mathematical description of the process of communication: a message elaborated by an information source that reaches the destination via a channel.It is clear that in this description both the concepts of states and transition between states are primitive objects and it is necessary to ask oneself what are their properties and what mathematical structure would properly encode them. From a more philosophical point of view, we could summarize the previous discussion by saying that a mathematical description of information sources, and more generally of communication processes, would require a way to implement both the concept of ``being'' and ``becoming''. It is, then, evident that these aspects will play a crucial role also when passing to the quantum description but with some differences, since there are plenty of experiments showing how distant quantum and classical mechanics could be. 

In this paper we are going to present some reflections on the concept of transition and how groupoids could actually encode their properties both from a classical and a quantum point of view. Even if some of these ideas have already been discussed in our previous works on the Schwinger's picture of Quantum Mechanics \cite{C-DC-I-M-02-2020,C-I-M-2018}, in this paper we are going to revisit those considerations from the point of view of a ``generalised logic''. Indeed, in a previous work \cite{C-DC-I-M-03-2020} the authors showed that the functions on a groupoid can be endowed with a von Neumann algebra \cite{Takesaki-2002} structure and the associated lattice of projections is an orthocomplemented lattice. This lattice, therefore, represents an example of a propositional calculus satisfying the axioms introduced by Birkhoff and von Neumann \cite{B-vN-1936}. But what can one say about transitions? The lattice approach, in this sense, seems to look at the logical structure of the experimental propositions providing information about the state of the system. How can it provide information about transitions? 

As a first attempt towards a statistical definition of transition, one could say that a transition between two events could happen if the occurrence of one could be conditioned by the other. Therefore, one could introduce a ``generalized logic'' whose basic propositions are pairs of ``experimental propositions'' that can be conditioned. In classical probability theory one implements the conditioning in a ``local'' way: the two events involved in the transition must have a non-empty intersection. Therefore, if we assume that the lattice of experimental proposition is atomic, the basic atoms cannot be conditioned, since they have empty intersections. However, one could go beyond this ``local'' approach introducing the concept of measurable groupoid. 

From an algebraic point of view, a groupoid is a set $\mathcal{G}$ of morphisms with a partial composition law, on which two maps are defined which associate to each morphism two objects, its source and target. These objects belong to another set, say $\Omega$. If we introduce a Boolean lattice of measurable subsets of the groupoid, we have a measurable groupoid. Using the composition law one can define a product among sets of the groupoid, whereas the source and the target maps send every set of morphisms to some sets of objects. Both these structures can be used to define a notion of ``non-local'' conditioning among subsets of the set of objects. In this paper we are going to show that using a measurable groupoid and introducing a generalized notion of conditioning between sets of objects, a measure on the groupoid determines a grade-2 measure on the set of objects \cite{Sorkin-1994}. Grade-2 measures where introduced by Sorkin in his formulation of Quantum Mechanics as a generalization of probability theory which could allow for the description of interference phenomena. Interestingly, using a groupoid the above mentioned grade-2 measure can be also associated with a state on the algebra of the functions on the groupoid and from this perspective the sets of zero 2-measure can be seen as elements in the Gelfand ideal of this state.  

\section{The lattice of propositions}
In this section we are going to shortly recall the basic notions introduced by Birkhoff and von Neumann to describe the logic of Quantum Mechanics, which is the set $\mathcal{P}$ of all ``experimental proposition'' that can be inferred about a system. These propositions are basic ingredients for the formulation of every physical model, and consequently they will share some common properties. In order to convey more clearly the idea we will restrict to a situation where only a finite set of propositions are considered. The basic relation existing on this set is an order relation, say $\subset$, which is the mathematical implementation of the implication relation between propositions. Once we have this order we assume that for every pair of elements $a,b \in \mathcal{P}$ there is a greatest lower bound, $a\wedge b$, and a least upper bound, $a\vee b$, which are, respectively, the greatest (with respect to the previous order) of all the propositions which are implied by $a$ and $b$, and the least of all the propositions which imply both $a$ and $b$. Additionally, we can always find two special propositions: the one which is always true, denoted by $\mathbb{I}$, and the one which is always false, say $\emptyset$. The final property of the set $\mathcal{P}$ is that for every proposition $a$ there is a complementary proposition $a^{\perp}$ satisfying the conditions $a\vee a^{\perp}=\mathbb{I}$ and $a\wedge a^{\perp} = \emptyset$. A set $\mathcal{P}$ endowed with the previous structures $\left(\subset,\,\wedge,\,\vee,\,\perp\,  \right)$ is called an orthocomplemented lattice, and this is the basic structure implementing the operations among ``experimental propositions''. The property that distinguishes the classical setting, however, is the distributive character of the two operations, i.e., $a\wedge (b\vee c) = (a\wedge b)\vee (a\wedge c)$. Then, when we have a finite distributive lattice $\mathcal{P}_c$ we can identify $\mathbb{I}$ with the set $\Gamma$ of all its atoms, and $\mathcal{P}_c$ will be isomorphic to the power set $\mathcal{P}(\Gamma)$. On the other hand, when we relax the distributive requirement, we can describe a lattice whose propositions are vector subspaces of a certain Hilbert space, which is the departure point in the von Neumann description of Quantum Mechanics. In particular, if one assumes a weaker form of the distributive identity, called modular identity, which in formulas reads
\begin{equation}
a\subset c\; \Rightarrow \; a \vee (b\wedge c) = (a \vee b) \wedge c\,
\end{equation} 
it is possible to prove that any complemented modular lattice of finite dimensions is isomorphic to the direct product of a finite Boolean algebra and a finite number of projective geometries. Moreover, if one additionally assumes irreducibility of the lattice, which means that the only elements $x\in \mathcal{P}$ satisfying 
\begin{equation}
a = (a\wedge x)\vee (a\wedge x^{\perp}) \;\;\forall a
\end{equation}
are $x=\emptyset,\,\mathbb{I}$, then the modular lattice is a finite projective geometry. In general by a projective geometry is meant any system satisfying the following axioms \cite{Birkhoff}: 
\begin{itemize}
\item Two distinct points are contained in one and only one line
\item If A, B, C are three points not on the same line and E$\neq$D are points such that B, C, D are on a line and A, C, E are on a line,then there is a point F such that both A, B, F are on a line and D, E, F are on a line
\item Every line contains at least three points
\item The points on lines through any k-dimensional element and a fixed point not in the element are a (k+1)-element, and every (k+1)-element is defined in this way.
\end{itemize}      
The set $\mathcal{C}(\mathcal{H})$ of Hilbert subspaces of the Hilbert space $\mathcal{H}$, is the prototype of a projective geometry. In this case k-dimensional elements correspond to (k+1)-dimensional closed subspaces of $\mathcal{H}$. On the other hand, one can replace any closed subspace with the associated orthogonal projector, obtaining a new lattice where the operation of meet and join, however, do not coincide with the sum and product of the algebra, generically. 

Once the structure of the experimental propositions is given, it is possible to associate to every event its probability: for Boolean lattices, monotone additive functions will play the role. For the modular case, in general, one can always define a dimension function $d\,\colon\,\mathcal{P}\,\rightarrow\,\left[0,1\right]$ obeying the following properties:
\begin{itemize}
\item $d(a)>d(b)\quad iff b\subset a$
\item $d(a) + d(b) = d(a\wedge b) + d(a \vee b)\,$.
\end{itemize} 
This dimension function corresponds to the normalized trace of each projection and represents the dimension of each closed subspace of the Hilbert space. More generally, however, one can consider real valued functions $m$ on the lattice which are additive only with respect to the meet of orthogonal projections. In this case Gleason theorem \cite{Gleason} ensures the existence of a density matrix $\rho$ such that $m(a) = \mathrm{Tr}(\rho P_a)$ with $P_a$ the projection corresponding to the closed subspace $a$ in the Hilbert space $\mathcal{H}$. This point of view, therefore, allows to give a statistical interpretation to each event, but only to events: we are interpreting only the ``static'' information. What about transitions? The standard approach introduced by von Neumann, define the probability transitions between two states of the system in terms of the square modulus of the scalar product of the vectors corresponding to the two states. In the rest of this short paper we will try to follow a different approach to the notion of transition.
\section{Groupoid and grade 2-measures}
In this section  we are going to introduce a generalized model of conditioning via the concept of groupoids. In order to avoid inessential technical difficulties and for the sake of communicability we are going to limit the discussion to finite groupoids and consider the extension to the non-finite case in forthcoming works. The departure point, therefore, is the definition of groupoids. From the algebraic point of view, a groupoid $G \rightrightarrows \Omega$, is a set $G$ of morphisms, together with a pair of maps $s,t\,\colon\,G\,\rightarrow\,\Omega$ from the set of morphisms to the set of objects $\Omega$, these maps being called source and target, respectively.  

Given the groupoid $G$, we will denote $G^j$ the set of morphisms whose target is $j\in \Omega$ (analogously we denote $G_j$ the set of morphisms whose source is $j$). The set of morphisms having the same source and target object $j\in \Omega$ is a group, the isotropy group at $j$, and will be denoted $G^j_j$. Two morphisms $\alpha$ and $\beta$ will be said to be composable if $s(\alpha) =t (\beta)$ and their composition will be denoted $\alpha \circ \beta$, this operation being associative. Units in the groupoid $G$ will be denoted as $1_j \colon j \to j$ and they satisfy the conditions
\begin{equation}
\alpha \circ 1_i = \alpha\,,\quad 1_j \circ \alpha = \alpha
\end{equation}
provided that $\alpha \colon i \to j$. Finally there is an inverse operation $\tau \colon \alpha \mapsto \alpha^{-1}$ such that 
\begin{equation}
\alpha^{-1} \circ \alpha = 1_i\,,\quad \alpha \circ \alpha^{-1} = 1_j\,.
\end{equation}

Some basic examples of groupoids, which will be used in the rest of the paper, are provided by the following ones. Firstly, the groupoid  of pairs $G(\Omega) = \Omega \times \Omega \rightrightarrows \Omega$ of an arbitrary set $\Omega$: it has source and target maps $s(j,i) = i$, $t(j,i) = j$, respectively, composition law $(k,j) \circ (j,i) = (k,i)$, units $1_i = (i,i)$ and inverse $(j,i)^{-1} = (i,j)$. Secondly, standard sets are groupoids, i.e., if $\Omega$ is a set we consider the groupoid $\Omega\,\rightrightarrows \Omega$ with only units (corresponding to each point of the set), so that a morphism is composable only with itself. 

Since a groupoid $G$ is a set of morphisms, we can consider the lattice $\mathcal{P}(G)$ of all its subsets and, as we have already discussed in the previous section, $\mathcal{P}(G)$ is a Boolean algebra. However, apart from this family of subsets, one can use the algebraic structure of the groupoid to generalize the ``local'' concept of conditioning expressed in terms of non-vanishing intersections. Indeed, if one considers two subsets $A,B \subset G$ it is possible to construct a new subset $C = B\circ A$ as follows:
\begin{equation}
C = \left\lbrace G\ni \gamma = \beta \circ \alpha\; \mid \; \beta\in B\,, \; \alpha\in A \right\rbrace\,.
\end{equation}  
In other words, $c$ is the set of all morphisms obtained by composing one morphism from the set $A$ and one from the set $B$. One can prove that this product is distributive with respect to the union of disjoint sets, but it is not commutative. In particular, one can take $A=s^{-1}(a)$ and $B=\tau^{-1}(s^{-1}(b)) = t^{-1}(b)$ with $a,b$ being two subsets of the set $\Omega$ of all objects of the groupoid. Then, we can say that two events $a,b \in \mathcal{P}(\Omega)$ are conditioned iff $\tau^{-1}(s^{-1}(b))\circ s^{-1}(a)\neq \emptyset$. It can be straightforwardly proven that conditioned pairs form the graphs of a reflexive and symmetric relation, which in general fails to be transitive. In the two main examples of groupoids above introduced, this definition can be specified as follows:
\begin{itemize}
\item for the groupoid $G(\Omega)\rightrightarrows \Omega$ of pairs of points of the set $\Omega$, one can consider only the products $(j,i)=\tau^{-1}(s^{-1}(j))\circ s^{-1}(i)$ where $i,j\in \Omega$ are atoms in $\mathcal{P}(\Omega)$, since the other sets are obtained via the distributive property of the product with respect to the union of disjoint events. It is then easily shown that $(j,i)=\left\lbrace(j,i)\right\rbrace$ which is the set containing the morphisms with source at $i$ and target at $j$.  
\item for the groupoid $\Omega\rightrightarrows \Omega$ made up of units only, the product $c = \tau^{-1}(s^{-1}(b))\circ s^{-1}(a) = b\wedge a$. 
\end{itemize} 

Once we have the lattice $\mathcal{P}(G)$ we can introduce a measure on $G$, say $\mu$. How can we use it to give a statistical interpretation to subsets of the groupoid? First of all, due to the algebraic structure of the groupoid, we will consider measures which are compatible with the composition and the inversion, in a way that generalizes the usual Haar measure for a group. Therefore, we will endow the groupoid $G$ with a measure $\mu$ which admits the following disintegration
\begin{equation}
\mu(A) = \sum_{j\in \Omega}\lambda(j) \nu^{j}(A)\,,
\end{equation}
with respect to a measure $\lambda$ on the space of objects $\Omega$. The family $\left\lbrace \nu^{j}\right\rbrace $ is a family of measures each one concentrated on the set $G^j$. It is called a left Haar system of measures if they satisfy the following properties:
\begin{itemize}
\item for $\alpha\colon i \rightarrow j$, $((L_{\alpha})_* \nu^{(i)}) = \nu^{(j)}$ where $(L_{\alpha})_*$ is the pushforward under the map $L_{\alpha}\,\colon\,G^i \,\rightarrow G^j$ representing the left multiplication by the morphism $\alpha$, i.e., $L_{\alpha}(\beta)= \alpha\circ \beta$. This property generalizes the left-invariance of the Haar measure to a framework where the composition among morphisms is only partially defined.
\item $\tau_{*}\mu = \delta^{-1}\mu$, where $\delta^{-1}$ is called the modular function and it is a homomorphism of the groupoid $G$ to the group of positive real numbers $\mathbb{R}_+$. This property means that by inversion we cannot transform a set of measure zero into a set of non-zero measure or viceversa. Once more, the modular function for groupoids generalizes the modular function for groups. 
\end{itemize}
The triple $(G,\mathcal{P}(G), \left[\mu\right] )$ with $\left[\mu\right]$ denoting the class of all measures equivalent to the left system of Haar measure $\mu$, will be called a measure groupoid. Let us remark that in this class, there is always one measure $\Lambda$ which is invariant with respect to the inversion, i.e., $\tau_*(\Lambda) = \Lambda$. Instead of considering the whole lattice $\mathcal{P}(G)$, let us focus, now, only on pairs of conditioned subsets $(b,a)$ in $\mathcal{P}(\Omega)\times \mathcal{P}(\Omega)$. Then, we can define the following two-set function $D\,\colon\,\mathcal{P}(\Omega)\times \mathcal{P}(\Omega)\,\rightarrow\,\mathbb{R}$:
\begin{equation}
D(b,a)= \Lambda (\tau^{-1}(s^{-1}(b))\circ s^{-1}(a))\,.
\end{equation} 
It can be easily seen, that this function satisfies the following properties:
\begin{itemize}
\item positivity $D(a,a)\geq 0$
\item bi-additivity $D(a,b\cup c)= D(a,b) + D(a,c)$ whenever $b\cap c = \emptyset$
\item symmetry $D(b,a) = D(a,b)$.
\end{itemize}
Let us remark that everything would work in the same manner if the measure $\Lambda$ would be weighted with a phase factor $e^{iS}$ with the function $S\,\colon\, G\, \rightarrow \mathbb{R}$ satisfying the ``logarithmic'' properties $S(\beta \circ \alpha)=S(\beta) + S(\alpha)$ and $S(\alpha^{-1})= -S(\alpha)$. The only difference would consists in the replacement of the symmetry condition with the hermiticity, since the function will be complex valued. 

The properties of the function $D$ characterize what is called a decoherence functional and consequently the associated quadratic form $\mu_2(a)=D(a,a)$ determine a grade 2-measure. Sorkin\cite{Sorkin-1994} introduced the notion of grade 2-measure in his generalization of probability theory in order to account for the description of quantum phenomena. Differently from a traditional measure, a grade 2-measure $\mu_2$ on a Boolean lattice $\mathcal{P}(\Omega)$ satisfies $\mu_2(a\vee b) - \mu_2(a) - \mu_2(b) = I(a,b)$
for every pair $(a,b)$ of disjoint subsets of $\Omega$. The two-set function $I(a,b)$ is the interference functional, and in principle does not vanish. However, a grade 2-measure satisfies the condition $0=I^{(3)}(a,b,c)=\mu_2(a\vee b \vee c) - \mu_2(a\vee b) - \mu_2(a\vee c) - \mu_2(b\vee c) + \mu_2(a) + \mu_2(b) + \mu_2(c)$.
Summarizing the above discussion, the algebraic structure of groupoids permit to introduce a generalized notion of conditioning between pairs of sets of a Boolean algebra. Once a suitable measure is introduced on the groupoid, a grade 2-measure is determined on the space of objects $\Omega$.
  
The above picture can be analyzed, also, from an algebraic point of view by considering the space of functions on the groupoid. Let $C(G)$ be the space of complex valued continuous functions on the finite groupoid $G$. On a finite groupoid this set is isomorphic to some vector space, say $\mathbb{C}^N$ where $N$ is the number of morphisms in $G$. Generalizing the construction for group-algebras, we can introduce a algebra structure on $C(G)$ as follows:
\begin{equation}
\begin{split}
f\star h = \left( \sum_{\alpha\in G} f_{\alpha} \delta_{\alpha} \right)\star \left( \sum_{\beta\in G} h_{\beta} \delta_{\beta} \right) = \\ = \sum_{\gamma\in G}\sum_{(\alpha,\beta)\in G^{(2)}\mid \alpha\circ\beta =\gamma}f_{\alpha}h_{\beta}\mu(\gamma)\delta_{\gamma}\,,
\end{split}
\end{equation} 
where $\delta_{\alpha}$ is the function which takes value 1 at $\alpha$ and 0 everywhere else, $\mu$ denotes the left invariant Haar system in the class $\left[ \mu \right]$, and $G^{(2)}$ denotes the set of composable pairs. This product, called convolution product, is associative. Moreover we can define the following involutive operator on $C(G)$:
\begin{equation}
f^{\dagger}=\sum_{\alpha \in G}\delta^{-1}(\alpha) \overline{f}_{\alpha} \delta_{\alpha^{-1}}\,,
\end{equation} 
so that $C(G)$ is endowed with a $\ast$-algebra structure. Here $\delta$ is the modular function associated with the measure groupoid structure. Moreover, it is possible to introduce a suitable topology\cite{I-R-2019} on this space so that we eventually get a von Neumann algebra, which we will denote $\nu(G)$. Once we have obtained this algebraic structure it is possible to express the previous decoherence functional in an alternative way. Indeed, it is easy to see that the product $A\circ B = C$ between subsets of $G$ can be expressed in terms of characteristic functions as follows:
\begin{equation}
\chi_C = \chi_A \star \chi_B\,,
\end{equation}  
where $\chi_C$ is the characteristic function supported on the set $C=A\circ B$, and analogously one defines $\chi_A$ and $\chi_B$. It can be proven \cite{CIM-2019} that the integration of a function $f\in \nu(G)$ with respect to the measure $\mu$ determines a positive linear functional on the von Neumann algebra which is normalized whenever the measure $\lambda$ on the set of objects $\Omega$ is a probability measure. Let us denote $\omega_{\mu}$ this state. Then, by direct computation one can show that 
\begin{equation}
D(b,a) = \omega_{\mu}((\chi_{s^{-1}(b)})^{\dagger}\star \chi_{s^{-1}(a)})\,.
\end{equation}  
Once this algebraic point of view is introduced, we can recognize that the sets whose corresponding characteristic functions are in the Gelfand ideal \cite{Takesaki-2002} of the state $\omega_{\mu}$ correspond to sets whose grade 2-measure $\mu_2$ is zero. Therefore, we can interpret the GNS construction as the construction of a suitable Hilbert space associated with the set of events of the groupoid.

\section{Conclusions}
In this paper we have presented some reflections on the notion of transitions, associated with the concept of pairs of conditioned events. After a brief review of the approach to quantum logic introduced by Birkhoff and von Neumann, we have discussed a different model where groupoids are the departure point. Indeed, using the rich algebraic structure defining a groupoid, we can introduce a way of conditioning sets which generalize the ``local'' approach of classical probability in a straightforward way. Of course, groupoid are not the only way to obtain this non-local conditioning. However, we have shown that using a groupoid this notion can be related to the $C^*$-algebraic approach to Quantum Mechanics and to the corresponding notion of state.    

\subsubsection*{Acknowledgements}FMC acknowledges that this work has been supported by the Madrid Government (Comunidad de Madrid-Spain) under the Multiannual Agreement with UC3M in the line of “Research Funds for Beatriz Galindo Fellowships” (C\&QIG-BG-CM-UC3M), and in the context of the V PRICIT (Regional Programme of Research and Technological Innovation). FDC thanks the UC3M, the European Commission through the Marie Sklodowska-Curie COFUND Action (H2020-MSCA-COFUND-2017- GA 801538) for their financial support through the CONEX-Plus Programme.
%
%
%
%

\end{document}